\documentclass[acmsmall,screen]{acmart}

\setcopyright{acmlicensed}
\copyrightyear{2024}
\acmYear{2024}
\acmDOI{XXXXXXX.XXXXXXX}

\acmISBN{978-1-4503-XXXX-X/18/06}

\usepackage{xspace}
\usepackage{listings}
\usepackage{xcolor}
\usepackage{multirow}
\usepackage{graphicx}
\usepackage{subcaption}

\begin{document}

\title{Identifying Root Causes of Null Pointer Exceptions with Logical Inferences}

\author{Jindae Kim}
\email{jindae.kim@seoultech.ac.kr}
\orcid{0000-0003-1864-0148}
\affiliation{%
  \institution{Seoul National University of Science and Technology}
  \city{Seoul}
  \country{South Korea}
}

\author{Jaewoo Song}
\email{jsongab@connect.ust.hk}
\orcid{0000-0003-0082-7967}
\affiliation{%
  \institution{The Hong Kong University of Science and Technology}
  \city{Hong Kong}
  \country{China}
}

\renewcommand{\shortauthors}{Kim and Song}
\newcommand{\logicfl}{\textsc{LogicFL}\xspace}
\newcommand{\smalltt}[1]{{\small\texttt{#1}}}
\newcommand{\pred}[1]{{\small\texttt{\textcolor{blue}{#1}}}}
\newcommand{\fig}{Fig.\xspace}
\newcommand{\tbl}{Table\xspace}
\newcommand{\lst}{Listing\xspace}
\newcommand{\prolog}[1]{{\small\texttt{\textcolor{blue}{#1}}}}
\newcommand{\java}[1]{{\small\texttt{\textcolor{purple}{#1}}}}

\lstdefinestyle{mystyle}{
    basicstyle=\ttfamily\small,
    breakatwhitespace=true,
    breaklines=true,                 
    captionpos=b,                    
    keepspaces=true,                 
    numbers=left,                    
    numbersep=5pt,                  
    showspaces=false,                
    showstringspaces=false,
    showtabs=false,                  
    tabsize=4,
    frame=lines,
    deletekeywords={name}
}

\lstset{style=mystyle}

\begin{abstract}
  Fault Localization (FL) is a task of locating a fault causing an abnormal behavior of software, and a crucial step of software debugging. To mitigate the burden of fault localization, many FL techniques have been proposed to automate FL tasks for a given test failure or an error. Recently, Large Language Model (LLM)-based FL techniques have been proposed, and showed improved performance with explanations on FL results.
  However, a major issue with LLM-based FL techniques is their heavy reliance on LLMs, which are often unreliable, expensive, and difficult to analyze or improve. When results are unsatisfactory, it is challenging both to determine a cause and to refine a technique for better outcomes.
  
  To address this issue, we propose \logicfl, a novel logical fault localization technique for Null Pointer Exceptions (NPEs). With logic programming, \logicfl imitates human developers' deduction process of fault localization, and identifies causes of NPEs after logical inferences on collected facts about faulty code and test execution. 
  In an empirical evaluation of 76 NPE bugs from Apache Commons projects and the Defects4J benchmark, \logicfl accurately identified the fault locations and pinpointed the exact code fragments causing the NPEs for 67 bugs (88.16\%), which were 19.64\% and 4.69\% more bugs than two compared LLM-based FL techniques respectively.
  Furthermore, for the bugs where fault locations were correctly identified, \logicfl found all the fault locations within the Top-3 and Top-5 candidates for 89.55\% and 98.51\% of the bugs respectively.
  In addition, \logicfl can be executed on a low-performance machine similar to a typical laptop, with an average runtime of 21.63 seconds and a worst-case time of under two minutes, including test execution and output file generation.
  Moreover, when compared to the two LLM-based FL techniques using the GPT-4o model, \logicfl was significantly more cost-efficient, as those techniques required 343.94 and 3,736.19 times the cost of \logicfl, respectively.
  Last but not least, the deduction process in \logicfl for providing FL results is fully traceable, enabling us to understand the reasoning behind the technique's outcomes and to further enhance the technique.
\end{abstract}



\maketitle

\section{Introduction}
\label{sec:intro}

Fault Localization (FL) is a task to locate a fault causing an abnormal behavior of a program.
Software debugging tasks take a significant amount of human developers' time~\cite{tassey2002economic}, and the majority of such debugging time is spent on bug diagnosis such as fault localization~\cite{debugging_time}.
To ease the burden, automated FL techniques have been proposed and studied, offering a variety of approaches that leverage diverse information and tools, tailored to different debugging scenarios~\cite{fl_survey1, fl_survey2}.

Recently, Large Language Model (LLM)-based FL techniques have been proposed, demonstrating impressive performance that surpasses previously existing methods~\cite{llm_fl, fusefl, LLMAO, autofl}.
These techniques also exploit LLMs' text generating ability to produce explanations about faults~\cite{autofl, fusefl}, which may address the weakness of automated FL techniques~\cite{fl_rationale1, fl_rationale2}.

However, a major issue with LLM-based FL techniques is ironically their heavy reliance on LLMs.
LLMs are often unreliable, expensive, and difficult to analyze or improve. 
Due to LLMs' probabilistic nature, LLM-based techniques need to be evaluated with repeated experiments~\cite{llm_fl}, or even embed the repetition to increase the confidence of a technique's outcomes~\cite{autofl}.
Moreover, even if LLMs can generate explanations for FL results, the rationale of generating such answers is still unknown, makes it difficult to address existing techniques' problems.

To address this issue, we propose \logicfl, a novel logical fault localization technique.
The key idea of \logicfl is imitating human developers' deduction process to identify the root cause of a given error.
It first collects logical facts from faulty code and test execution, then applies defined rules which represent the knowledge regarding fault localization and programming languages.
Since a final outcome is deduced based on defined rules and collected facts, we can understand the rationale behind the outcome, and also analyze or improve the deduction process using our knowledge.

To prove that this direction is possible and effective, we selected Null Pointer Exceptions (NPEs) as our first target.
Although the NPE looks quite simple to be addressed, it has been revealed frequent errors~\cite{npe_significance, android_crash_vitals}, and many FL techniques~\cite{npe_fl_1, npe_fl_2, npe_fl_3} and Automatic Program Repair (APR) techniques~\cite{nopol, PAR, npe_repair_1, NPEFix} have taken NPEs into consideration.
Despite the significance of the problem, our knowledge to identify NPE causes is relatively simple and well-organized, hence make it suitable for \logicfl.

\logicfl's logical rules are designed to imitate common deduction process to find a cause of an NPE.
First, \logicfl identifies a null expression which made an exception thrown.
Then starting from the null expression, it identifies culprits which may transfer null to that expression. 
During the deduction, \logicfl uses collected facts containing information of source code, test failures, stack traces and observed values of expressions, just like human developers would do while debugging. 
With this information, it applies rules to trace how null value is transferred.

The defined rules represent our knowledge on NPEs and common behaviours of programs. 
For instance, we know that an NPE may happen if an expression is referenced while it is null. 
\logicfl searches for expressions which satisfy these conditions to identify null expressions.
Also, we know that if a return statement of a method returns null, then an expression calling the method may have null value.
Such knowledge is embedded as logical rules, and used during deduction process.

For empirical evaluation, we compared \logicfl with two state-of-the-art LLM-based FL techniques, FuseFL and AutoFL. 
We applied \logicfl and the other techniques to 76 NPE bugs collected from four Apache Commons projects (38 bugs) and Defects4J benchmark (38 bugs).
Then we measured how many bugs' fault locations can be matched with Top-10 candidates.
\logicfl correctly identified all fault locations for 67 (88.16\%) of the bugs, which were 19.64\% and 4.69\% more bugs than the matched bugs of AutoFL and FuseFL respectively.
Furthermore, for the bugs whose fault locations were correctly identified, \logicfl found all the fault locations within Top-3 and Top-5 candidates for 89.55\% and 98.51\% of the cases respectively. 
In addition, we evaluated runtime performance and costs of \logicfl, with a low-performance machine similar to typical laptops.
\logicfl processed a bug in average 21.63 seconds, and the total estimated cost of \logicfl to process all 76 bugs was only \$0.0153.
For the same bugs, AutoFL and FuseFL's costs for using the GPT-4o model were \$57.1637 and \$5.2623, which were 3,736.19 and 343.94 times of the \logicfl's cost respectively.

We also analyzed how \logicfl's rules contributed to identify fault locations and rank candidates.
We found that all of the defined rules contributed to identify at least one fault location.
However, the rules to rank candidates were not very effective for some cases.
This result demonstrates that \logicfl's deduction process is fully traceable, allowing us to understand the reasoning behind the technique's outcomes and further improve it.

Here is the summary of our contributions:

\begin{itemize}
\item We propose \logicfl, a logical fault localization technique for NPEs. To the best of our knowledge, this is the first attempt to combine static and dynamic analysis with rules written in a logic programming for fault localization.
\item We provide a benchmark of 76 NPE bugs for fault localization, including identified fault locations and NPE causes.
\item We conducted empirical evaluation of \logicfl to show the effectiveness and efficiency of \logicfl.
We also provide empirical evaluation results of FuseFL and AutoFL with a more recent LLM as well as estimated cost, which were not provided in the previous studies~\cite{fusefl, autofl}.
\end{itemize}

The rest of the paper is organized as follows. 
We first provide background of this study (Section~\ref{sec:back}) and explain detailed approach of \logicfl in Section~\ref{sec:logicfl}.
Sections~\ref{sec:eval} and~\ref{sec:results} present experimental settings and empirical evaluation results respectively.
Sections~\ref{sec:discussion} and~\ref{sec:threats} further discuss \logicfl's capability with future directions and threats to validity respectively, then we conclude the study in Section~\ref{sec:conclusion}.
\section{Background}
\label{sec:back}

We provide background for fault localization as well as logic programming, which is necessary to understand \logicfl's core idea.

\subsection{Fault Localization}

To mitigate the burden of locating faults, many automated FL techniques have been proposed.
Although these techniques are different in details, they often share common characteristics and can be categorized to several prevalent categories~\cite{fl_survey1,fl_survey2}.
Among various FL approaches, Spectrum-based FL (SBFL) techniques are the one which showed the best standalone FL performance~\cite{fl_survey2}, and there are frequent SBFL techniques~\cite{ochiai, tarantula, dstar, op2, barniel} which have been studied and compared repeatedly and showed reliable FL performance~\cite{sbfl_empirical,sbfl_empirical2, sbfl_empirical3}.

Some other attempts leveraged new information or analysis results such as static analysis~\cite{Neelofar}, crash stacks~\cite{CrashLocator}, suspicious variables~\cite{vfl}, code and change metrics~\cite{fluccs}, semantics and probabilistic model~\cite{SmartFL}, explainable artificial intelligence techniques~\cite{xai4fl}, causal inference~\cite{UniVal}, and many others to improve SBFL performance.
Also, Machine Learning (ML) techniques have been also frequently adopted to form a new trend of ML-based FL (MLFL) techniques~\cite{deepfl, GRACE, DeepRL4FL}.

More recently, a new direction of utilizing LLMs has been explored, showing promising results that surpass many existing FL techniques.
Wu et al. investigated the FL performance of ChatGPT-4, and found that it outperformed all the existing compared FL techniques~\cite{llm_fl}.
Yang et al. proposed LLMAO, which employed a left-to-right language model with a bidirectional adaptor to train classifier to predict fault locations, and showed that LLMAO can outperform compared MLFL techniques~\cite{LLMAO}.
FuseFL utilized Chain of Thought (CoT)~\cite{CoT1, CoT2}, test results, suspiciousness score and code descriptions in prompts of LLMs to improve FL performance, as well as generating comprehensive explanations for its output~\cite{fusefl}.
AutoFL addresses the issue of limited prompt size by providing tools which can retrieve necessary information from code base for LLMs, showed comparable or superior FL performance compared to existing FL techniques~\cite{autofl}.

The advancements brought by LLMs in FL are impressive and encouraging; however, we have several concerns regarding LLM-based FL techniques. 
First, LLM-based methods might be unreliable. 
It is not just the inconsistency of outcomes, but also the difficulty in understanding how they are generated.
Moreover, training and operating LLMs demand significant resources and costs~\cite{stanford_ai_report_2024}, and access to such models might be limited due to this reason.
Therefore LLM-based FL techniques often discussed a better way to use LLMs~\cite{llm_fl, fusefl, autofl} rather than improving a model behind it~\cite{LLMAO}.

To address these issues, we propose \logicfl, which mimics human's deduction process of fault localization. 
The most important resource to implement \logicfl is our knowledge to define rules for fault localization.
Since fault localization process is actually logical inferences based on the defined rules, this automated process is fully traceable and logically understandable.

\subsection{Logic Programming}

To imitate human's deduction for fault localization, \logicfl leverages a logic programming language, Prolog.
Logic programming languages are often used for knowledge representation and reasoning~\cite{KRR}.
Since we try to embed our knowledge to deduce causes of NPEs in \logicfl, using logic programming is a reasonable choice.
More specifically, we used SWI-Prolog~\cite{swiprolog} to implement \logicfl, but we will use Prolog for simplicity throughout the paper.
We will briefly explain how Prolog works, since it is necessary to understand how \logicfl's rules work.

\begin{lstlisting}[language=Prolog, caption={An Example Knowledge Base}, label={lst:kb_1}]
parent(jill, jane). % Facts.
parent(john, jane).
male(john).
female(jill).
father(jack, jill).
father(X, Y) :- parent(X, Y), male(X). % Rules.
grandfather(X, Y) :- father(X, Z), parent(Z, Y).
\end{lstlisting}

\lst~\ref{lst:kb_1} is an example of \emph{Knowledge Base (KB)} written in Prolog.
The first five predicates are \emph{facts}, which literally indicate known facts, and facts are true by themselves.
For example, the first predicate indicates that \prolog{jill} is a parent of \prolog{jane}. 
Both \prolog{jill} and \prolog{jane} are \emph{atoms}, which start with a lowercase letter, and used to represent constants or names. 
We can also refer to a specific type of predicate using \smalltt{<predicate name>/<arity>}, such as \pred{parent/2} predicates in the first two lines.

Lines 6 and 7 contain \emph{rules}, which is of the form \smalltt{<head> :- <body>}.
Predicate in the head and body parts are also called \emph{goal} and \emph{sub-goals} respectively.
A goal is true only if all the sub-goals are true.
For instance, \prolog{father(X, Y)} indicates that \prolog{X} is a father of \prolog{Y}.
The first rule defines that \prolog{father(X, Y)} is true, if \prolog{X} is a parent of \prolog{Y} and \prolog{X} is male.
In this rule, \prolog{X} and \prolog{Y} are variables which start with an uppercase letter, and can be instantiated with a certain value later.
Note that a comma (,) means \emph{Logical AND}, and a semi-colon (;) represents \emph{Logical OR} in Prolog.
A period (.) indicates the end of a fact or rule, and these facts and rules end with a period are called \emph{terms}.

Once a KB is constructed, we can use it by issuing a \emph{query}.
For instance, consider a query \prolog{?- father(X, Y).}
The question is whether there exist \prolog{X} and \prolog{Y} that \prolog{X} is the father of \prolog{Y}.
Prolog searches the KB from \emph{top to bottom}, to find facts or rules which can be unified with the given query.
Simply speaking, two terms can be unified if we make them identical strings, by instantiating variables with values.
The first term which can be matched with the query is \prolog{father(jack, jill)} at line 5.
For \prolog{father(X, Y) = father(jack, jill)}, we have a solution \prolog{X = jack, Y = jill}, which makes both side of the equation identical.
The next term can be matched with the query is the first rule \prolog{father(X, Y)} at line 6.
To satisfy a rule, Prolog tries sub-goals from \emph{left to right}.
In this case, \prolog{parent(X, Y)} can be unified with the first two facts, but only \prolog{parent(john, jane)} can also satisfy \prolog{male(john)}.
Hence it produces another solution \prolog{X = john, Y = jane}.

Remind that Prolog only provides answers based on the facts and rules in a given KB.
For instance, based on our common knowledge, we can deduce that \prolog{jill} is the mother of \prolog{jane}.
However, for a query \prolog{?- mother(jill, jane)}, Prolog will answer \prolog{false}, since there is no specified fact or rule for \pred{mother/2}.
To identify fault locations and causes of NPEs, \logicfl constructs a KB consisting of collected facts and defined rules, so that Prolog can correctly deduce answers based on the facts.
\section{Logical Fault Localization}
\label{sec:logicfl}

In this section, we explain \logicfl, a \textbf{Logic}al \textbf{F}ault \textbf{L}ocalization leveraging logical inferences to identify causes of NPEs.

\subsection{Overview}

\begin{figure}[thb]
    \centering
    \includegraphics[width=0.95\linewidth]{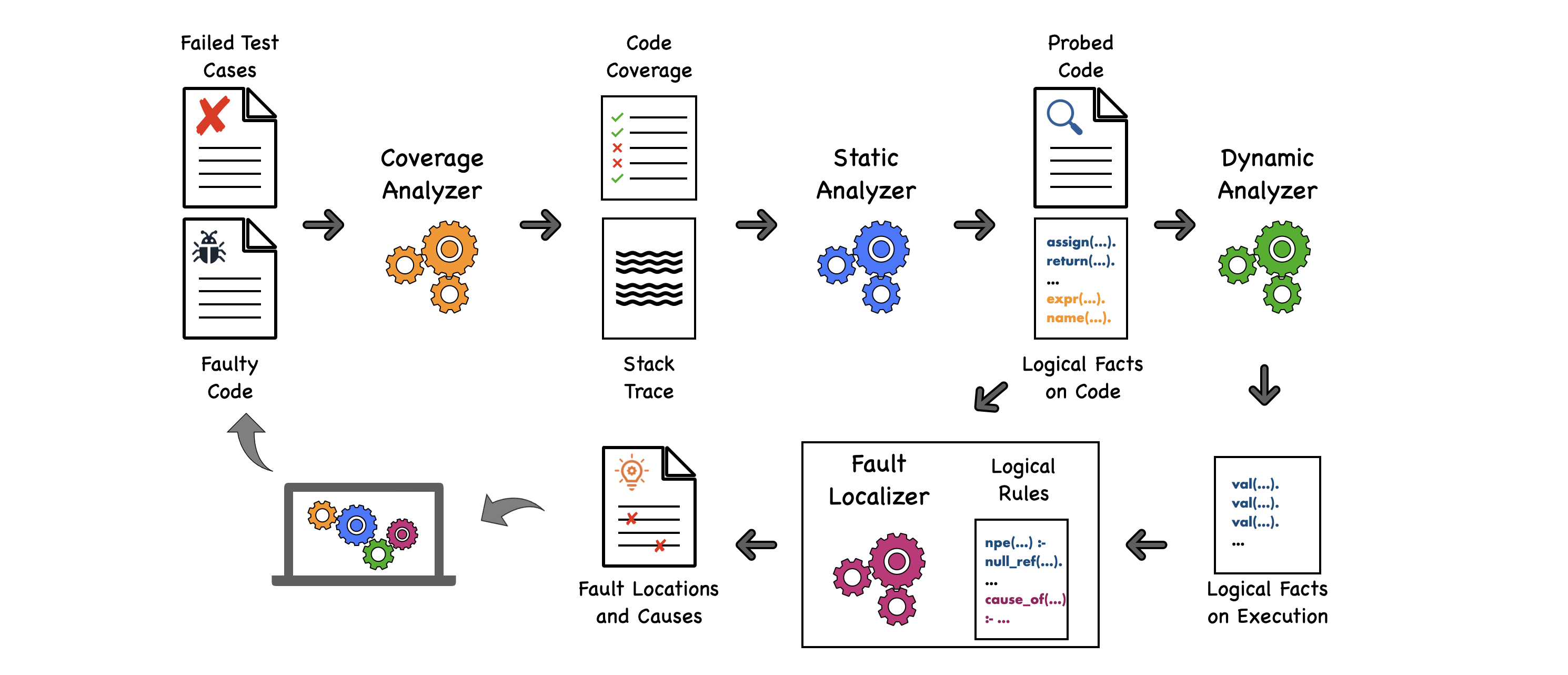}        
    \caption{Logical Fault Localization Process for Null Pointer Exceptions}
    \Description{The figure describes an overview of LogicFL. It shows the inputs and outputs produced by each component of LogicFL, as well as their connection during the fault localization process.}
    \label{fig:overview}
\end{figure}

\fig~\ref{fig:overview} shows the overall process of \logicfl.
First, it takes faulty code and failed test cases reproducing NPEs as inputs. 
Note that \logicfl requires at least one failed test to start with, but it can also operate with multiple test failures.
The first step of \logicfl is applying \emph{CoverageAnalyzer}, to obtain code coverage and stack traces.
Then \emph{StaticAnalyzer} uses this information to collect logical facts from covered code lines.
Another important role of StaticAnalyzer is producing \emph{Probed Code} for instrumentation.
\emph{DynamicAnalyzer} executes failed tests on the probed code, and monitors actual values of expressions.
Its primary role is generating facts on test execution, which contains information of observed expression values.
Finally, all the collected facts are given to \emph{FaultLocalizer}.
FaultLocalizer applies logical rules to the facts, and provides fault locations and NPE causes deduced by the rules.

\subsection{Collecting Logical Facts}

One of the key considerations in fact collection is \emph{traceability} of code entities, which can be achieved by assigning the same name to the same code entity.
In Prolog, these code entities are appeared as atom, and atoms' identity is decided by their names.
Hence assigning the same name allows \logicfl to easily recognize the same code entities.

Consider a code fragment `\java{return stream(stream.stream());}' from one of the NPE bugs.
The first and the third \smalltt{stream} are methods with different parameters, and the second \smalltt{stream} is a variable.
Also, there are two method invocations, one is \java{stream.stream()} and the other is \java{stream(stream.stream())}.
Finally, the code fragment is a return statement, returning the outermost method invocation.

For this code fragment, StaticAnalyzer generates two kinds of facts: Semantic facts and Code Facts. Listing~\ref{lst:static_facts} shows an example of collected semantic and code facts.

\begin{lstlisting}[language=Prolog, caption={An Example of Semantic and Code Facts}, label={lst:static_facts}]
method_invoc(expr2, m_stream_1, line(streams_1, 479)). % Semantic facts
ref(p_stream_1, expr2, line(streams_1, 479)).
method_invoc(expr1, m_stream_2, line(streams_1, 479)).
argument(expr2, 1, expr1).
return(expr1, m_stream_3, line(streams_1, 479)).

class(streams_1, 'org.apache.commons.lang3.stream.Streams'). % Code facts
expr(expr2, m_invoc, expr1, (args, 0), range(streams_1, 19452, 15, 479, 479), "stream.stream()").
name(p_stream_1, simple_name, expr2, expr, range(streams_1, 19452, 6, 479, 479), 'stream').
\end{lstlisting}

\emph{Semantic facts} represent the meaning we understand about the code.
The first fact in Listing~\ref{lst:static_facts} is a \pred{method\_invoc/2} predicate representing \java{stream.stream()}. 
This fact indicates that \prolog{expr2} is a method invocation of the method \prolog{m\_stream\_1}, at line 479 of a class \prolog{streams\_1}.
Similarly, the second fact shows that \prolog{p\_stream\_1} (\java{stream}) is referenced in \prolog{expr2}.
Also, \prolog{argument(expr2, 1, expr1)} indicates that \prolog{expr2} is the first argument of another method invocation \prolog{expr1}.
Lastly, \prolog{return(expr1, m\_stream\_3, line(streams\_1, 479))} shows that \prolog{expr1} is returned in another \java{stream} method \prolog{m\_stream\_3}, which is different from \prolog{expr1} and \prolog{expr2}'s methods.

\emph{Code facts} provide references for atoms used in semantic facts, and store code structure to connect them with source code.
The first code fact at line 7 shows the qualified name of the \prolog{streams\_1} class.
The other \pred{expr/6} and \pred{name/6} predicates represent expressions and names.
The fact at line 8 shows various information, such that \prolog{expr2} is a method invocation (\prolog{m\_invoc}), which is also an argument (\prolog{(args, 0)}) of \prolog{expr1} with index 0.
It also shows the exact code range with a \pred{range/5} predicate, as well as the actual code.

Another important role of StaticAnalyzer is generating \emph{probed code} by injecting probes which can monitor expressions' values. Listing~\ref{lst:probed_code} shows an example probed code after StaticAnalyzer injected probes to the return statement.

\begin{lstlisting}[language=Java, caption={An Example of Probed Code}, label={lst:probed_code}]
Collection<O> p_stream_1_line_479 = stream;
Stream<O> expr2_line_479 = (Stream<O>) (p_stream_1_line_479.stream());
FailableStream<O> expr1_line_479 = stream(expr2_line_479);
return streams_1_expr1_line_479;
\end{lstlisting}

A \emph{probes} is a variable declaration containing an expression to be monitored as an initializer.
For instance, line 1 in Listing~\ref{lst:probed_code} contains a probe \java{p\_stream\_1\_line\_479} for a parameter \java{stream}.
Probe injection maintains the order of original program execution.
Once a probe is injected for an expression, the other part of code uses the probe's name instead, so that the expression is evaluated only once.
Each probe's name contains the name of an atom assigned to the expression (\java{p\_stream\_1}) as well as its line number, hence DynamicAnalyzer can correctly generate facts of observed values for the atom.

How DynamicAnalyzer works is similar to a developer monitors test execution using debugging support.
It first adds break points to code lines appeared in NPE stack traces, then records the values of declared variables in probes.
Currently it only focuses on expressions with null values and generates \pred{val/3} predicates as facts.
For instance, suppose that the parameter \smalltt{stream} at line 1 is null during test execution. 
DynamicAnalyzer generates a fact \prolog{val(p\_stream\_1, null, line(streams\_1, 479))}, using the information from the probe's name \prolog{p\_stream\_1\_line\_479}.

Collected semantic and code facts provide necessary information for deduction of NPE causes.
Using such facts, \logicfl understands how faulty code works, and deduce which part of the code could be a cause of an NPE.
Due to the limitation of space, we did not include all types of defined facts, but a complete specification of defined facts can be found in our replication package~\cite{replication}.

\subsection{Identifying Causes}

Once all facts are collected for an NPE bug, these facts are combined with pre-defined rules to form a KB for the given NPE bug.
We defined 73 rules in total, including 25 rules for identifying NPE causes and 48 rules for understanding common behaviour of programs.
Due to page limits, we will present only the key rules for identifying NPE causes; the complete set of the rules is publicly available~\cite{swish}.

Listing~\ref{lst:find_npe_cause} shows \pred{find\_npe\_cause/4}, which is used as the main query to identify NPE causes.
With a constructed KB of an NPE bug, FaultLocalizer issues the query \prolog{?- find\_npe\_cause(Expr, Line, Cause, Loc)} to obtain answers about NPE causes.

\begin{lstlisting}[language=Prolog, caption={The \pred{find\_npe\_cause/4} Rule for the Main Query}, label={lst:find_npe_cause}]
find_npe_cause(Expr, Line, Cause, Loc) :- 
    findall((Expr, Line, Cause, Loc), 
        (npe(Expr, Line), cause_of(npe(Expr, Line), Cause, Loc)), 
        TotalCauses),
    remove_duplicates(TotalCauses, UniqueCauses),
    rank_causes(UniqueCauses, RankedCauses),
    member((Expr, Line, Cause, Loc), RankedCauses).
\end{lstlisting}

The key part of this rule is two sub-goals \prolog{npe(Expr, Line)} and \prolog{cause\_of(npe(Expr, Line), Cause, Loc)} at line 3.
First, \prolog{npe(Expr, Line)} identifies a null expression \prolog{Expr} at \prolog{Line}, which directly threw an NPE.
This is the starting point of tracking the actual cause of the NPE.
With the implementation of JEP 358 in Java 14~\cite{JEP358}, null expressions are given in error messages, hence human developers can easily identify them in source code.
\pred{npe/2} rule is designed to provide the same information with collected facts about code and stack traces.
After that, \prolog{cause\_of(npe(Expr, Line), Cause, Loc)} finds the actual cause \prolog{Cause} and its location \prolog{Loc} for a null expression \prolog{Expr} at line \prolog{Line}.

One NPE bug (faulty code) may have more than one test failures, hence multiple pairs of \prolog{(Expr, Line)} can be found.
\logicfl first identifies all possible causes in \prolog{TotalCauses}, using \pred{findall/3} at line 2-4. 
Then it removes duplicates with \pred{remove\_duplicates/2} at line 5.
For the \prolog{UniqueCauses}, \pred{rank\_causes/2} checks preferred conditions and filtering conditions, to rank up or filter out each candidate respectively.
Finally, \pred{member/2} provides each candidate one by one in the order stored in \prolog{RankedCauses}.

Listing~\ref{lst:cause_of} provides five rules for \pred{cause\_of/3}.
These rules are describing six combinations of two null expression types and three identification schemes. 
First, \logicfl identifies \emph{Null Ref.} and \emph{Null Arg.} type null expressions as starting points.
Then it applies three identification schemes, \emph{Direct}, \emph{Origin}, and \emph{Transfer}, to identify a root cause from the null expressions.

\begin{lstlisting}[language=Prolog, float, caption={Rules to Identify NPE Causes and Their Locations}, label={lst:cause_of}]
cause_of(npe(Expr, Line), Cause, Loc) :- 
    null_arg_passed(Expr, Line),
    Cause = Expr, Loc = Line.
cause_of(npe(Expr, Line), Cause, Loc) :- 
    null_arg_passed(Expr, Line),
    originated_from(val(Expr, null, Line), (Expr1, Line1)),
    Cause = Expr1, Loc = Line1.
cause_of(npe(Expr, Line), Cause, Loc) :-
    null_ref(Expr, Line),
    Cause = Expr, Loc = Line.
cause_of(npe(Expr, Line), Cause, Loc) :-
    null_ref(Expr, Line),
    originated_from(val(Expr, null, Line), (Expr1, Line1)),    
    Cause = Expr1, Loc = Line1.
cause_of(npe(Expr, Line), Cause, Loc) :-
    (null_arg_passed(Expr, Line) ; null_ref(Expr, Line)),
    can_be_transferred(val(Expr, null, Line), (Cause, Loc)).
\end{lstlisting}

In each rule, the first sub-goal represents null expression types.
\pred{null\_ref/2} and \pred{null\_arg\_passed/2} are the rules to find Null Ref. and Null Arg. type null expression respectively.
Null Ref. type indicates that a referenced expression is null, like \java{stream} in \java{stream.stream()}.
Null Arg. type is a case that null is passed as an argument to a method, like \java{path} in \java{Files.isSymbolicLink(path)}.
In this case, not passing null argument might be necessary, since we cannot modify standard library \java{isSymbolicLink()}.

Three identification schemes are implemented by the remainder of sub-goals in each rule.
In Direct scheme, the null expression itself is considered as a cause, like lines 3 and 10 of Listing~\ref{lst:cause_of}.
We often address an NPE by adding a null checker right before a null expression, hence considering the null expression as a cause is a practical choice.
Origin scheme traces back to the origin of null value transferred to the null expression (\pred{originated\_from/2} at lines 6 and 13).
For null safety, it is also a good solution to use a default instance instead of null, hence the origin of null which does not pass such an instance can be considered as a cause.
Transfer scheme enumerates all possible locations which may transfer null to the null expression (\pred{can\_be\_transferred/2} at line 17).
This scheme provides candidates between the null expression and the origin, who act as conduits for null value transfer, which may need to be interrupted to resolve the issue.

Since Prolog searches a KB from top to bottom, the order of the rules in Listing~\ref{lst:cause_of} is the order that \logicfl considers each combination and provides candidates.
\logicfl tries Direct and Origin schemes with Null Arg. and Null Ref. first.
Then it applies Transfer scheme regardless of null expression types.
We chose this order since it was effective in our preliminary investigation during rule development, but the order can be adjusted based on further analysis results.
A more important point is that we can control the order of candidates.

To trace back to the origin and identify expressions along the way, \logicfl uses various rules to understand common behaviour of code.
\pred{copied\_from/2} rules shown in the following are an example of such rules.

\begin{lstlisting}[language=Prolog, numbers=none]
copied_from((Expr, Line), (Expr1, Line)) :- assign(Expr, Expr1, Line).
copied_from((Expr, Line), (Expr1, Line1)) :- method_invoc(Expr, M, Line), 
                                             return(Expr1, M, Line1).
\end{lstlisting}

In the first rule, \prolog{assign(Expr, Expr1, Line)} represents an assignment \prolog{Expr = Expr1} at line \prolog{Line}. 
In this case, we know that \prolog{Expr}'s value is copied from \prolog{Expr1}'s value.
Similarly, the second rule shows a case of a return statement and a method invocation.
If \prolog{Expr1} is returned in a method \prolog{M}, a method invocation \prolog{Expr}'s value calling \prolog{M} at \prolog{Line} is copied from \prolog{Expr1}'s value at \prolog{Line1}.
There are many other rules like them to represent common behaviour of programs.
With these rules, we represent our knowledge and grant abilities to \logicfl to deduce programs' behaviour.

After candidates are identified, \pred{rank\_causes/2} rule at line 6 of Listing~\ref{lst:find_npe_cause} is applied to rank up preferred candidates and filter out undesirable candidates using the rules in Listing~\ref{lst:rank}.

\begin{lstlisting}[language=Prolog, caption={Rules to Rank Candidates}, label={lst:rank}]
prefer_cond((_, _, Cause, Loc)) :- 
    find_method(Method, Loc), 
    (is_null_return(Method, Cause) ;
    	val_assigned_in_method(Cause, Loc, Method) ; 
    	only_target_method(Method)), !.
filter_cond((_, _, _, Loc)) :- from_test(Loc), !.
filter_cond((_, _, _, Loc)) :- find_method(Method, Loc), arg_passing_method(Method), !.
\end{lstlisting}

\logicfl considers three preferred conditions, \pred{is\_null\_return/2}, \pred{val\_assigned\_in\_method/3}, and \pred{only\_target\_method/1} (lines 3-5), and filters out candidates by checking two filtering conditions, \pred{from\_test/1} and \pred{arg\_passing\_method/1} (lines 6-7).
Note that an underscore (`\_') represents a wildcard, and it was used to mask information that we do not need in these rules.

The preferred conditions describe the conditions of preferred candidates.
\pred{is\_null\_return/2} checks whether a candidate is a null literal returned in a method, while the method also returns something other than null.
For such cases, using an empty or default instance instead of null might be desirable, hence we set this as a preferred candidate - or a recommended fix location.
\pred{val\_assigned\_in\_method/3} indicates that there is an assignment to a candidate inside the same method it is located. 
In this case, this candidate might be a better choice compared to the other methods which do not change candidates' values.
\pred{only\_target\_method/1} verifies that a given method is the only non-test method in the code base appeared in stack traces.
Compared to other candidates executed, but not appeared in stack traces, these candidates are likely to have more strong relation to NPEs.

The filtering conditions define undesirable characteristics of candidates.
\pred{from\_test/1} simply checks if a candidate is from a test, which likely to introduce null for test execution, but not in our concerns.
\pred{arg\_passing\_method/1} verifies if a candidate is in a method simply passing arguments to another method (i.e., only has a return statement calling another method).
In this case, it would be better to handle a problem in the callee, which might contain more computation in its body.


\logicfl utilizes these rules with collected facts, and imitates human's deduction process to identify NPE causes.
This process is fully traceable using Prolog's \pred{trace/1} predicate, which provides a step-by-step view of the logical inferences. 
Furthermore, after constructing a KB from a given NPE bug, we can query the KB for analysis and define new rules to infer additional information.
\section{Experimental Settings}
\label{sec:eval}

In this section, we explain how we conducted empirical evaluation of \logicfl including Research Questions (RQs), an evaluation dataset and details of experimental settings.

\subsection{Research Questions}

\paragraph{RQ1. \textbf{How many bugs does \logicfl accurately identify NPE causes?}}
\logicfl applies the rules to the collected facts to identify NPE causes. 
Although the rules are designed based on common knowledge about why NPE happens, it is necessary to evaluate whether they are effective to correctly identify NPE causes.
Hence we applied \logicfl to NPE bugs and counted the number of the bugs which it correctly identified the fault locations and causes.

\paragraph{RQ2. \textbf{How well does \logicfl identify NPE causes?}}
Even if \logicfl can identify NPE causes correctly, it is not helpful if developers need to go through many possible candidates. 
Hence for each bug, we considered at most 10 candidates recommended by \logicfl, and measured how quickly the actual fault locations and causes can be found in the list of candidates.

\paragraph{RQ3. \textbf{How efficient is \logicfl to identify NPE causes?}}
The efficiency of an FL technique is important if it is to be used practically during software development.
We evaluated \logicfl's runtime performance and costs by executing it on a cloud instance with low-performance similar to typical laptops.

\paragraph{RQ4. \textbf{How effective are the rules to identify NPE causes?}}
Since \logicfl applies rules for fault localization, understanding the effectiveness of each rule is important to understand how it works, as well as to improve and refine the current set of rules.
We traced deduction process of \logicfl, and analyzed the usage of the rules to evaluate their effectiveness in FL tasks.

\subsection{Null Pointer Exception Benchmark}
\label{sec:benchmark}

\begin{table}[htb]
    \centering
    \small
    \caption{Statistics of the Null Pointer Exception Benchmark}
    \begin{tabular}{l|cccc||l|c}
    \hline
        Project & \#Bugs & \#FailedTests & \#FaultLocs & \#Causes & Bug Types & \#Bugs \\
    \hline
       collections  & 3 & 13 & 11 & 11 & Single Test & 48 \\       
       commons-io & 19 & 29 & 31 & 32 & Multiple Tests & 28 \\
       lang & 13 & 17 & 16 & 16 & Single FaultLoc & 53\\       
       math  & 3 & 7 & 5 & 5 & Multiple FaultLocs & 23 \\
       defects4j & 38 & 90 & 53 & 54 & Single Cause & 52 \\
    \cline{1-5}
       Total & 76 & 156 & 116 & 118 & Multiple Causes & 24 \\
    \hline
    \end{tabular}    
    \label{tbl:sbj}
\end{table}

We created an NPE benchmark consists of 76 NPE bugs to evaluate \logicfl.
We collected 38 bugs from four Apache Commons projects and 38 bugs from Defects4J 2.0.1 benchmark~\cite{defects4j_github}.
The Apache Commons projects were chosen for their reliability in reproducing NPEs. 
Also, Defects4J benchmark has been widely used in many FL and APR related studies~\cite{llm_fl, autofl, LLMAO, LLM_APR, AlphaRepair}.

Apache Commons NPE bugs were collected from four projects: Lang, Math, Collections, and IO.
We examined all commits in their development history up to November 2, 2023 for the first three, and February 23, 2024 for IO.
We searched for commits which mentioned ``null'' or ``NPE'' (case insensitive) that included bug fixes and test cases.
Then we collected buggy and fixed versions as well as necessary information for evaluation, such as failed tests and error messages.

After collecting the NPE bugs, we identified their causes by manual inspection considering the following points.
Firstly, we checked developer's patches to identify code fragments with issues.
For example, if a patch inserted a null checker \smalltt{var != null}, we considered \smalltt{var} as a prime candidate of the cause.
Secondly, we verified that these candidates indeed produced NPEs of failed tests.
For instance, in \smalltt{var.method1().method2()}, if \smalltt{var.method1()} is null but \smalltt{var} is not null, we only consider \smalltt{var.method1()} as a cause.
Also, there were some fix locations that corresponding failed tests were not available.
We also excluded these cases, as they did not actually cause NPEs but were modified as preemptive fixes.
We recorded these as potential faults, but labeled them as `fixed' rather than `faulty'.
This manual inspection process was conducted by one of the authors and reviewed by three independent auditors.

The statistics of the NPE benchmark is shown in \tbl~\ref{tbl:sbj}.
On the left side, we report the number of bugs (\#Bugs), failed tests (\#FailedTests), identified fault locations (\#FaultLocs) and causes (\#Causes).
In the benchmark, a \emph{cause} is defined as the code fragment that needs to be considered to remove an NPE, while a \emph{fault location} is the specific line of code containing that cause.
We represented a cause in a certain class by its code range (i.e., starting index and length), and fault location by its line number.
On the right, we also report the number of bugs in different categories. 
\emph{Multiple Tests} refers to bugs with more than one failed test, and \emph{Multiple FaultLocs} indicates bugs with multiple fault locations. 
\emph{Multiple Causes} represents bugs with more than one cause. 
There are 28 bugs with multiple tests, but only 23 have multiple fault locations, meaning some tests failed due to the same NPE cause at the same location. 
Additionally, one more bug falls under Multiple Causes than Multiple FaultLocs, as two causes at the same location triggered different test failures.

\subsection{Decisions on Matching of Fault Locations and Causes}
\label{sec:fl_decision}

To measure FL accuracy of \logicfl, we need to decide whether a bug's fault locations or causes are correctly identified.
We counted the number of actual fault locations and NPE causes appeared in Top-10 candidates of an FL technique, and categorized them to \emph{matched}, \emph{partially matched}, and \emph{not matched} cases.
Note that we considered Top-10 candidates since there are two bugs with seven fault locations in the NPE benchmark.
If all fault locations of a bug are identified, the bug is matched. 
If a technique identifies at least one of the locations, but not all of them, then the bug is partially matched. 
The bug is classified as not matched if none of the locations are found. 
We apply the same policy for NPE causes.

\subsection{Compared FL Techniques}
\label{sec:compared}

For empirical evaluation, we compared \logicfl with two state-of-the-art FL techniques leveraging LLMs: AutoFL~\cite{autofl} and FuseFL~\cite{fusefl}.
To automate evaluation process as much as possible, we used OpenAI's API (Application Programming Interface) service for LLM-based techniques.
AutoFL provides its implementation to automatically use API and obtain FL results, but FuseFL does not provide such implementation.
We implemented a simple tool to automatically collect necessary information and generate a FuseFL prompt for each bug to obtain results.

We also needed to adjust a few settings for AutoFL and FuseFL.
For AutoFL, we modified its prompt and computation of results to support line-level fault localization.
AutoFL was developed to produce method-level fault locations~\cite{autofl}. 
Originally, it asked to ``provide the signatures of the most likely culprit methods," but we rephrased this part to ``suggest which lines would be the best locations to be fixed," and asked for exact locations in \smalltt{ClassName.MethodName:LineNumber} format.

FuseFL requires faulty code, code description, test results, and SBFL results to construct its prompts.
Unlike FuseFL's original evaluation with students' assignments, we could not insert the entire code base as faulty code, hence we provided code lines executed by failed tests as faulty code.
For code descriptions, we generated code descriptions using an LLM by following the suggested method in FuseFL~\cite{fusefl}.
For test results, we provided test code and error messages.
Finally, we decided to remove SBFL part.
The original experiments did not reveal which formula was used to compute SBFL results, it was not possible to exactly replicate the original prompt~\cite{fusefl}.
It also did not show too much degradation in performance without SBFL results, hence we decided to remove this part from the prompts.
Moreover, with failed tests only, most of the code lines would have the same SBFL scores.
Other than this, we tried our best to follow the original prompt.

In addition, we accounted for the fact that both AutoFL and FuseFL rely on probabilistic models, which can cause variations in FL performance. 
To ensure fairness, we conducted 10 trials for these two techniques.
For FuseFL, we selected the trial with the best performance for analysis.
AutoFL, on the other hand, has a feature that combines results from repeated runs to produce a final outcome. 
Therefore, we also performed 10 runs for AutoFL and used the combined results.
Note that the default configuration for the number of runs ($R$) in AutoFL is 5, and the authors of AutoFL reported degradation of performance with $R> 5$~\cite{autofl}.
In our settings, we found that using $R = 10$ produced more improved results (14\% more matched bugs) than $R = 5$, hence we stayed with 10 runs.

\subsection{Execution Environment}

For runtime performance and cost estimation, we need to control the execution environment of \logicfl.
For this purpose, we executed \logicfl on Amazon EC2 t3.small instance with 2 vCPU (virtual Central Processing Unit) and 2 GiB memory.
Since each vCPU indicates a thread of a CPU core~\cite{amazon_ec2_cpu_options}, the performance of the machine is similart to typical laptops.
We chose this type of instance intentionally, because we considered a scenario that \logicfl is used locally by individual software developers during their software debugging tasks. 

For LLM-based FL techniques, we used OpenAI's API service with the GPT-4o model.
GPT-4o was one the most advanced models available~\cite{openai_models} during the evaluation, and FuseFL and AutoFL have not been tested on this model.
More specifically, at the time of experiment, GPT-4o model pointed to \smalltt{gpt-4o-2024-05-13} model.
The benefits from utilizing more advanced models is one of the strengths of LLM-based FL techniques.
Therefore we decided that it was worth to verify the LLM-based FL techniques' performance with a more recent model.

\section{Results}
\label{sec:results}

In this section, we present and discuss about empirical evaluation results of \logicfl.

\subsection{RQ1. Fault Localization Accuracy}

\begin{table}[H]
\centering
\small
\caption{Fault Location Identification Results on the NPE Benchmark. The numbers of bugs and ratios to total 76 bugs are shown for Matched, Partially Matched and Not Matched categories.}
\begin{tabular}{l|cccc}
\hline
\textbf{Apache Commons} & \textbf{\logicfl} &\textbf{AutoFL} & \textbf{FuseFL} & \textbf{AutoFL+FuseFL}\\
\hline
Matched & 35 (92.11\%) & 28 (73.68\%) & 35 (92.11\%) & 36 (94.74\%) \\
Partially Matched & 3 (7.89\%) & 6 (15.79\%) & 2 (5.26\%) & 2 (5.26\%) \\
Not Matched & 0 (0.00\%) & 4 (10.53\%) & 1 (2.63\%) & 0 (0.00\%) \\
\hline
\textbf{Defects4J} & \textbf{\logicfl} & \textbf{AutoFL} & \textbf{FuseFL} & \textbf{AutoFL+FuseFL}\\
\hline
Matched & 32 (84.21\%) & 28 (73.68\%) & 29 (76.32\%) & 32 (84.21\%) \\
Partially Matched & 0 (0.00\%) & 3 (7.89\%) & 1 (2.63\%) & 1 (2.63\%) \\
Not Matched & 6 (15.79\%) & 7 (18.42\%) & 8 (21.05\%) & 5 (13.16\%) \\
\hline
\textbf{Total} & \textbf{\logicfl} & \textbf{AutoFL} & \textbf{FuseFL} & \textbf{AutoFL+FuseFL}\\
\hline
Matched & 67 (88.16\%) & 56 (73.68\%) & 64 (84.21\%) & 68 (89.47\%) \\
Partially Matched & 3 (3.95\%) & 9 (11.84\%) & 3 (3.95\%) & 3 (3.95\%) \\
Not Matched & 6 (7.89\%) & 11 (14.47\%) & 9 (11.84\%) & 5 (6.58\%) \\
\hline
\end{tabular}
\label{tbl:matched}
\end{table}

\tbl~\ref{tbl:matched} shows the fault localization results of \logicfl, AutoFL and FuseFL.
The numbers of Matched, Partially Matched and Not Matched bugs and ratios are shown for Apache Commons NPE bugs and Defects4J NPE bugs, as well as the combined Total results.
AutoFL+FuseFL column provides fault localization results when we consider the best outcome of the two techniques (i.e., we counted a better outcome of AutoFL and FuseFL for each bug).

Overall, \logicfl outperforms AutoFL and FuseFL in all aspects.
\logicfl correctly identified all fault locations of 67 bugs, which is 88.16\% of the bugs.
AutoFL and FuseFL have 56 (73.68\%) and 64 (84.21\%) matched bugs respectively.
In terms of fully matched bugs, \logicfl has 19.64\% and 4.69\% more matched bugs than AutoFL and FuseFL, respectively.
FuseFL showed FL performance more close to \logicfl, but this was the best result among 10 trials.
The number of matched bugs varied from 61 to 64, and there were only two trials with 64 matched bugs.
We reported the result of a trial which had more matched bugs with Top-1 candidate, but still \logicfl had better overall FL performance.
\logicfl also had only six bugs whose fault locations were not matched, fewer than the 11 (AutoFL) and 9 (FuseFL) bugs that were not matched by the other techniques.
Even if we consider two sets of NPE bugs from Apache Commons and Defects4J separately, still \logicfl outperforms the other FL techniques in both sets in terms of more matched bugs and fewer not matched bugs.
Note that we also verified matched bugs with NPE cause results, and the numbers of bugs were identical to fault location results in \tbl~\ref{tbl:matched}.

Moreover, \logicfl's FL performance is even comparable to the combined performance of AutoFL+FuseFL.
If we aggregate all matched bugs of them, there are 68 matched bugs and 5 not matched bugs, which is slightly better outcome than \logicfl.
Even with a substantial advantage, the combined LLM-based technique shows slightly better performance than \logicfl.
The LLM-based techniques used the same model, hence the capability of the model itself was unchanged.
However, how we used this model affected the FL results, and provided different outcomes for the same problems presented by the two techniques.
Since \logicfl relies on logical rules representing our knowledge on fault localization, it consistently provides reliable performance. 
On the other hand, the LLM-based techniques missed some bugs when used individually, but this was not due to capabilities of the underlying model, as the AutoFL+FuseFL results suggest. 

\subsection{RQ2. Fault Localization Efficiency}
\label{sec:rq2_result}

\begin{figure}[htb]
    \centering
    \includegraphics[width=0.98\textwidth]{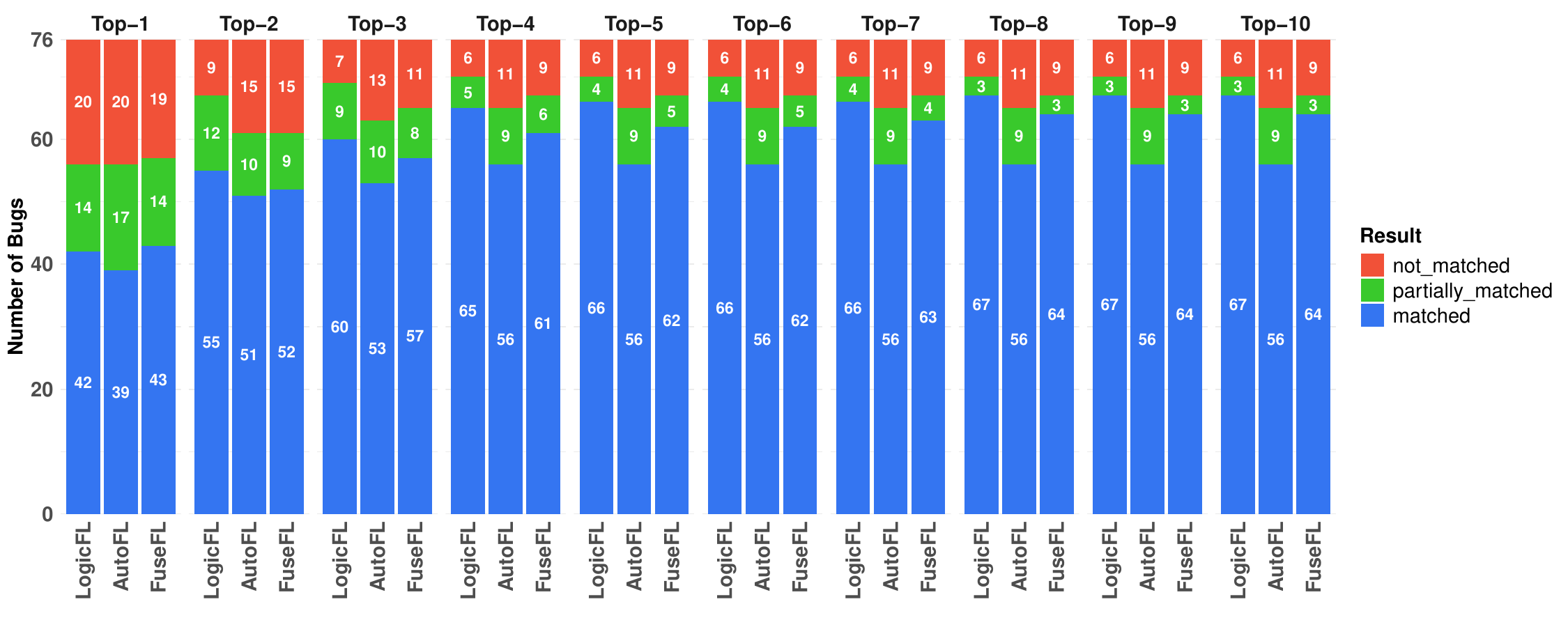}
    \caption{Fault Localization Results of Top-10 Candidates.}
    \Description{This chart shows the number of bugs whose fault locations are matched, partially matched or not matched when only TopN candidates are considered.}
    \label{fig:top10}
\end{figure}

\fig~\ref{fig:top10} shows Top-10 matched results of the three FL techniques. 
Bottom blue bars indicate the number of matched bugs when only Top-N candidates are considered. 
Green bars in the middle show partially matched bugs, and red bars at the top represent not matched bugs for Top-N candidates.
For each Top-N candidates, \logicfl, AutoFL, and FuseFL's results are shown from left to right.

Except for Top-1, \logicfl continuously outperforms AutoFL and FuseFL.
For Top-1, FuseFL has one more matched bug and one fewer not matched bug than \logicfl.
After that, we can see \logicfl's blue bars are the highest from Top-2 to Top-10, and the red bars of \logicfl are the shortest too.
\logicfl already has 60 matched bugs at Top-3, which are 89.55\% of all matched bugs, and after considering Top-5 candidates, \logicfl found 98.51\% of the matched bugs.
There is only one bug matched with the 8-th candidate of \logicfl, and this bug has seven fault locations.
AutoFL and FuseFL also identifies fault locations efficiently, with all matched bugs found within the Top-4 and Top-8 candidates respectively.

To further analyze efficiency, we computed Average Unnecessary Examination (AUE) for matched bugs.
AUE is the average of differences between the number of examined candidates and actual fault locations of matched bugs.
With AUE, we can estimate how much unnecessary effort was needed until finding all the fault locations for a bug.


Among the three techniques, AutoFL showed the lowest AUE, and \logicfl had the highest AUE.
The AUE values of \logicfl, FuseFL and AutoFL are 0.36, 0.33, and 0.21 respectively.
This indicates that all three techniques require fewer than one additional candidate examinations on average.
If a bug can be matched, AutoFL requires the least effort, which confirms the Top-N result that AutoFL found all matched bugs within Top-4, while others found in Top-8.
We applied Kruskal-Wallis rank sum test to unnecessary examinations of the three techniques.
The result with p-value 0.364 (>0.05) suggests that the difference among the three techniques is not statistically significant.

\subsection{RQ3. Runtime Performance and Cost}
\label{sec:runtime}

\begin{figure}[htb]
    \centering
    \includegraphics[width=0.8\linewidth]{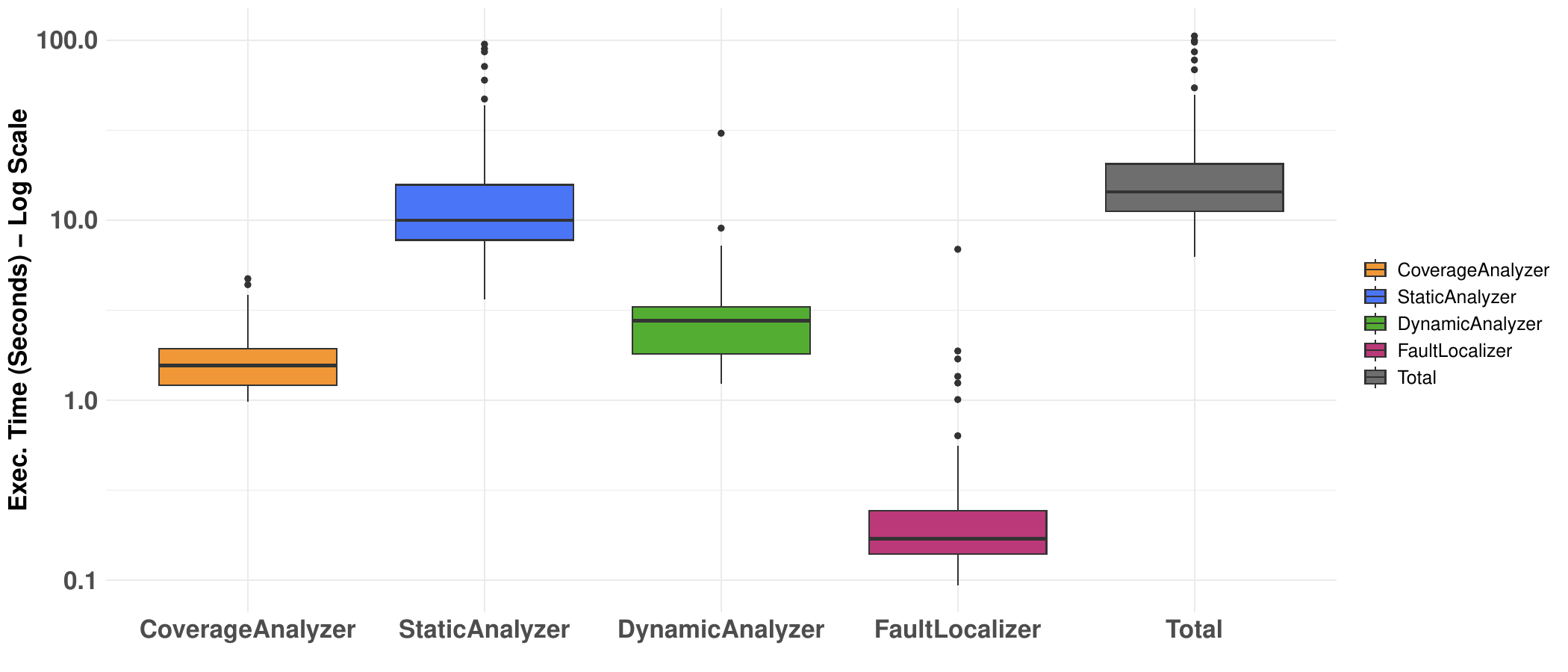}
    \caption{Execution Time of \logicfl Components Measured on Amazon EC2 t3.small}
    \Description{This chart shows the execution time of LogicFL in seconds.}
    \label{fig:exec_time}
\end{figure}

\fig~\ref{fig:exec_time} presents box plots of execution time distribution for each component as well as the entire \logicfl.
Since there exists huge difference among the components' execution times, we used log scale for the plots.
Outliers are shown as dots in the plots, and the median values are represented as a thick bar within the boxes.

The average total execution time of \logicfl was 21.63 seconds.
This means that \logicfl can be run on a laptop-level device within 22 seconds on average.
The maximum total execution time of \logicfl was 105.40 seconds, hence it took under two minutes even for the worst case.
Since we measured the entire execution time including reading and writing files, 22 seconds is the average execution time we can expect if we use \logicfl in practice.

Each component contributes differently to the total execution time of \logicfl, with StaticAnalyzer consuming the majority, averaging 70.56\%. 
CoverageAnalyzer and DynamicAnalyzer account for 10.37\% and 17.67\% of the time respectively, while FaultLocalizer requires only 1.40\%.
StaticAnalyzer collects logical facts from all covered Java classes, hence it is understandable that this component takes the most of the time to process many classes.
If we employed a strategy to process only modified classes, StaticAnalyzer would work more faster after facts of all classes are collected.
Note that FaultLocalizer applies defined rules in 0.37 seconds on average, 6.89 seconds even for the worst case.
Therefore we may add more rules without increasing execution time significantly, to improve \logicfl's performance.

\begin{table}[thb]
\centering
\caption{Fault Localization Cost of \logicfl, FuseFL and AutoFL. `Ratio to \logicfl' rows show the total cost of FuseFL and AutoFL relative to that of \logicfl.}
\begin{tabular}{l|ccc}
\hline
Fault Localization Cost & \textbf{\logicfl} & \textbf{AutoFL} &  \textbf{FuseFL}\\
\hline
Total Cost - GPT-4o (US\$) & 0.0153 & 57.1637 & 5.2623 \\
Total Cost - GPT-4o-mini (US\$) & 0.0153 & 1.7796 & 0.1665\\
\hline
Ratio to \logicfl (GPT-4o) &  \textbf{1.00} & \textbf{3,736.19} & \textbf{343.94}\\
Ratio to \logicfl (GPT-4o mini) &  \textbf{1.00} & \textbf{116.32} & \textbf{10.89}\\
\hline
\end{tabular}
\label{tbl:cost}
\end{table}

We also estimated the cost of \logicfl, AutoFL and FuseFL.
For \logicfl, we measured the cost of using Amazon EC2 t3.small instance, based on the total execution time.
The hourly charge varies based on the regions~\cite{amazon_ec2_pricing} around \$0.02-0.03, and we applied the highest price \$0.0336 for \logicfl.
For AutoFL and FuseFL, we measured the number of input and output tokens, and multiplied unit prices of a model~\cite{openai_api_pricing}.
GPT-4o model requires \$5 and \$15 for 1M input and output tokens respectively, and GPT-4o-mini has unit prices \$0.15 and \$0.6 for 1M input and ouput tokens respectively.

\tbl~\ref{tbl:cost} presents the estimated costs for \logicfl, AutoFL and FuseFL across 76 bugs.
\logicfl's cost was only \$0.0153, while AutoFL and FuseFL required significantly higher amounts at \$57.1637 and \$5.2623, respectively.
These are 3,736.19 (AutoFL) and 343.94 (FuseFL) times of the cost spent by \logicfl, which indicate that \logicfl is significantly more cost-efficient.
On average, \logicfl processed a bug with only \$0.0002.
Note that FuseFL required 9.21\% of AutoFL's cost, since we only report the cost of the best trial, not all 10 repeated trials, while AutoFL combined all 10 trials. 
We also did not include costs for generating code descriptions.

We also calculated the estimated costs using a more cost-efficient GPT-4o-mini model, by multiplying its unit prices to measured tokens.
Even with this model, AutoFL and FuseFL still require \$1.7796 and \$0.1665, which are 116.31 and 10.89 times the cost of \logicfl, respectively.
LLMs can serve more wide ranges of tasks, but that does not reduce their costs when we use them for a specific task.
Despite its extremely low cost, \logicfl consistently outperforms the other LLM-based techniques in fault localization performance.

\subsection{RQ4. Effectiveness of the Rules}

\begin{figure}[htb]
    \centering
    \begin{subfigure}[b]{0.45\textwidth}
        \centering
        \includegraphics[width=\textwidth]{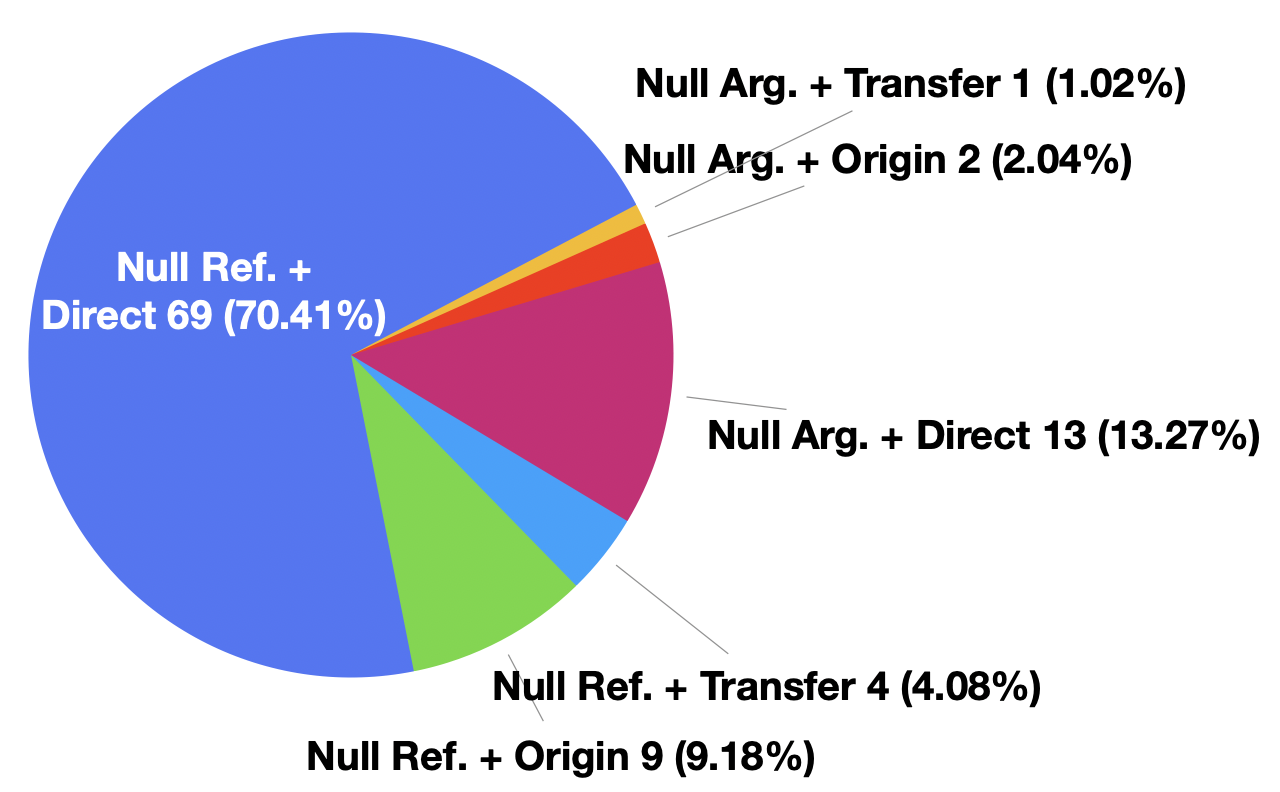}
        \caption{Usages of Rules to Identify Fault Locations}
        \label{fig:npe_rules}
    \end{subfigure}
    \hfill
    \begin{subfigure}[b]{0.45\textwidth}
        \centering
        \includegraphics[width=\textwidth]{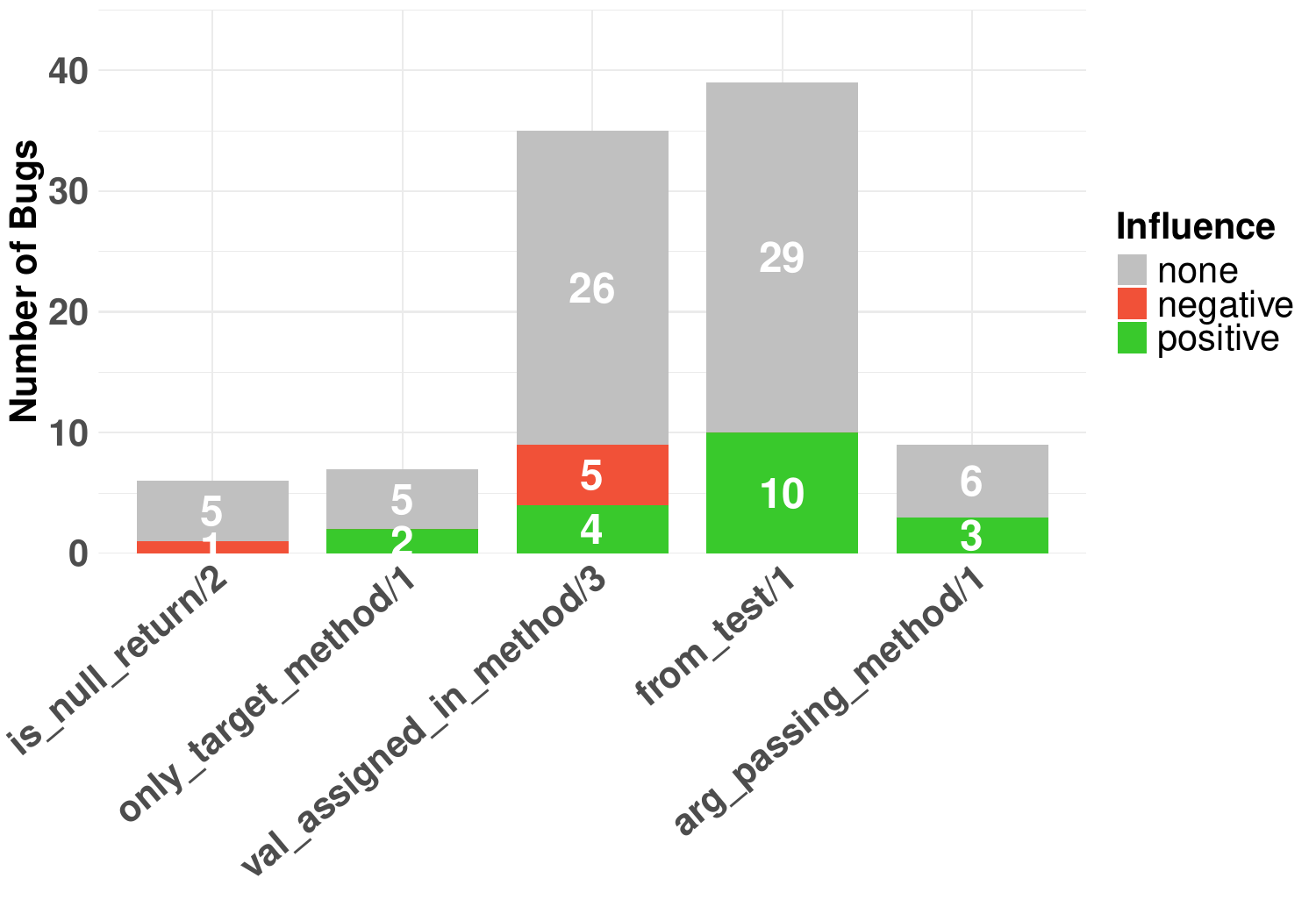}
        \caption{Influence of Rules to Rank Candidates}
        \label{fig:rank_rules}
    \end{subfigure}
    \caption{Rule Usages and Influence of \logicfl on the NPE Benchmark.}
    \Description{Usages of rules to identify fault locations are shown as a pie chart. Influence of rules to rank candidates is represented with stacked bar charts, for each of the five rules for preferred conditions and filtering conditions.}
    \label{fig:rules}
\end{figure}

We analyzed effectiveness of the rules to identify NPE causes (Listing~\ref{lst:cause_of}) and rank candidates (Listing~\ref{lst:rank}).
First, we counted the number of times the six combinations in Listing~\ref{lst:cause_of} were used to identify fault locations. 
Then we investigated the influence of the rules in Listing~\ref{lst:rank}.

\fig~\ref{fig:npe_rules} shows a pie chart drawn with the number of fault locations identified by each pair of null expression type and identification scheme.
Null Ref. type and Direct scheme pair is most frequent, which accounts for 70.41\% of the all identified fault locations.
Next frequently used pair is Null Arg. and Direct, which takes 13.27\%.
Hence 83.68\% of the fault locations were identified directly from the locations which NPEs were thrown.
The other four combinations were also responsible for at least one identified fault location from 16.32\% of the fault locations, which indicates that every rule contributed for fault localization.

There are 11 fault locations which were identified by Origin scheme.
These fault locations belong to 10 NPE bugs, and unlike \logicfl, either FuseFL or AutoFL failed to match 5 out of 10 bugs.
Since we provided covered code lines, the origin itself must be included in them, but identifying the origin as a cause was not successful for LLM-based techniques.
However, analyzing why LLM-based techniques provide such undesirable outcome would be difficult, if we only had information obtained from LLM-based techniques themselves.
On the other hand, we can fully monitor deduction process of \logicfl, and analyze the roles of each defined rule.
Comparison to \logicfl even provides insights to recognize weakness of LLM-based techniques in FL results.

\fig~\ref{fig:rank_rules} shows the influence of the five rules used to rank candidates.
We assessed their influence as \emph{none}, \emph{positive}, and \emph{negative}.
If a rule is applied but does not change the rank of any actual fault locations, the influence is none.
A positive influence increases the rank of a fault location, while a negative influence decreases it.
We counted the number of bugs having candidates which satisfied these rules, and categorized their influence.

Overall, filtering rules were helpful to increase FL performance, but preferred rules were not very effective.
The two filtering predicates had no negative influence, yet showed positive influence for 13 (10 for \pred{from\_test/1} and 3 for \pred{arg\_passing\_method/1}) bugs. 
On the other hand, two of the three preferred rules had negative influence on ranking candidates.
\pred{is\_null\_return/2} had no positive influence, while showed negative influence on one bug.
\pred{val\_assigned\_in\_method/3} had both positive and negative influence, but there was one more bug having negative influence.
\pred{only\_target\_method/1} was the only predicate which showed positive influence without any negative effect.
These results indicate that current rules to select preferred candidates are not very effective, and this might be the reason why \logicfl was less efficient than the others (Section~\ref{sec:rq2_result}).
We can use this analysis result to improve the performance of \logicfl, by revising preferred conditions or adding more filtering conditions.
\section{Discussion}
\label{sec:discussion}

In this section, we further discuss about \logicfl's capability and future directions.

\subsection{Not Matched Bugs}

As shown in \tbl~\ref{tbl:matched}, \logicfl failed to identify fault locations of six NPE bugs.
All of these bugs are from Defects4J benchmark: Chart-2, Gson-9, Math-70, Math-79, Jsoup-66, and Closure-171.
These bugs have fault locations which cannot be identified by considering null transfers.


For instance, consider a method from Gson-9 bug: \java{public JsonWriter value(boolean value)}.
An NPE happens when a test calls \java{value((Boolean) null)}. 
Since there was no \java{value(Boolean)} method, the call was directed to the above method taking a primitive \java{boolean} type parameter.
During this process, an implicitly type conversion from a wrapper type to a primitive type occurred, resulted in the NPE due to the type conversion of null.
The bug was fixed by adding a new \java{value(Boolean)} method. We identified the type \java{boolean} of the old method as the cause, since it is necessary to be considered to fix this problem.
\logicfl does not have rules to recognize such implicit conversion at the moment, hence it failed to identify the cause.
The other five bugs also have similar issues, requiring undefined rules to express missing common behaviors or complex domain knowledge.

To identify such causes, we need to add more rules to handle exceptional cases like these bugs.
Defining such rules may require significant effort and would only address a limited number of cases. 
Therefore, we must carefully balance the effort and the potential benefits when deciding whether to add new rules.

An interesting observation is that, of the six bugs not matched by \logicfl, the first three bugs were fully matched by AutoFL or FuseFL. 
While \logicfl excels at identifying NPE causes using general knowledge, these exceptions, which require more specific insights, may be better addressed by LLM-based FL techniques trained on a wide range of cases. 
Unless we can find a way to automatically deduce logical rules for handling such exceptions, logic-based approaches will inherently struggle with these outliers.

\subsection{Extension to Other Error Types}

While \logicfl demonstrates better FL performance than LLM-based techniques, the latter have advantages in terms of generality. 
LLMs process source code and other information like error messages or code coverage as plain text, allowing them to be applied to various types of errors, issues, and programming languages — though their effectiveness may vary.
On the other hand, \logicfl requires a clearly defined knowledge base including specific types of facts and rules designed to represent our knowledge about a certain problem.
Currently, we only showed that it works for identifying causes of NPEs.

However, this does not imply that \logicfl lacks the capability of extension.
With StaticAnalyzer, we can already collect wide ranges of logical facts about source code.
Based on these facts, we also provide set of rules to describe some common behaviour of programs.
DynamicAnalyzer can be further extended to collect a wider variety of facts about program execution, not just the values of expressions that are null.

Based on these facts and rules, we can design a new set of rules to provide solutions for other issues such as different error types or security vulnerability.
As we discussed in Section~\ref{sec:runtime}, adding more rules will not increase the execution time of \logicfl significantly.
Therefore, for problems that we have clear, well-organized knowledge, the approach designed for \logicfl may offer a highly reliable and efficient automated solution.



\subsection{Generation of Explanation}
\label{sec:explanation}

One of the key issues addressed by LLM-based FL techniques is providing explanations alongside identified fault locations. 
FuseFL generated correct explanations for 22 out of 30 cases~\cite{fusefl}, while AutoFL provided accurate descriptions for 56.7\% of the bugs in question~\cite{autofl}.
\logicfl only provide specific locations of faulty code fragments, and it does not generate comprehensive natural language descriptions on its outcomes.
Although \logicfl has clear rationale for every decision it makes, it is not easy to produce natural, comprehensive descriptions to explain all the process.

However, this does not mean that explanations for \logicfl's fault localization results cannot be provided. 
Unlike other existing techniques, each identified candidate fault location of \logicfl is backed by reasoning.
What we need to do is simply find a way to generate easily understandable descriptions of this reasoning.
Additionally, to provide accurate explanations for why a candidate is the correct fault location, we must first ensure that it is indeed the correct location. 
Given \logicfl's impressive accuracy in fault localization, there is potential for it to generate accurate descriptions for a larger number of bugs.

Moreover, providing natural language descriptions might not be crucial in different contexts.
For instance, many APR techniques do not require natural language description~\cite{ssFix, SimFix, PAR, confix, sharpFix, SapFix}, rather they prefer correct fault locations.
We can also consider an extension of an Integrated Development Environment (IDE), jumping to each code fragment appeared on null transfers during \logicfl's deduction process with a simple tag for reasoning.
In this way, we can still provide proper rationale to help debugging, without generating detailed descriptions in a long text.


\section{Threats to Validity}
\label{sec:threats}

There are several concerns about internal validity.
First, our judgment on identified NPE causes in the NPE benchmark might be incorrect and subjective.
To avoid potential issues, we tried our best to systematically decide NPE causes based on defined conditions, and multiple human evaluators verified identified causes (Section~\ref{sec:benchmark}).
Also, the changes we applied for FuseFL and AutoFL might affect the evaluation results, despite of our best effort to follow the original approaches as much as possible.
These two techniques assumed different conditions to operate, hence still our results show their performance in different situations.
Lastly, data leakage issue that the training data of LLMs may include our dataset, could affect AutoFL and FuseFL's performance.
Although this may affect the validity of the reported results, it favored the compared techniques, without changing the conclusion that \logicfl outperformed them.

The external validity threat is mostly due to the limited NPE benchmark.
We evaluated \logicfl's performance on 76 NPE bugs from open source projects.
However, these bugs may not fully represent the broader population of NPE bugs, and as a result, \logicfl's performance could vary on a different dataset.
To mitigate this limitation, we collected all identified NPE bugs from the entire software history of four projects, ensuring a more representative sample of the overall population of NPE bugs.
Additionally, we included all NPE bugs from a widely-used benchmark, enabling continuous comparison and testing with other techniques.

The construct validity threat arises from how we measured FL performance. 
We primarily used the number of matched bugs as the key indicator of FL performance and considered the number of matched bugs within the Top-N candidates to assess the efficiency of each technique. 
However, these measurements may not fully capture the practical usefulness of the techniques in real-world FL tasks. 
Specifically, unlike FuseFL and AutoFL, \logicfl does not provide natural language descriptions for identified fault locations, meaning that relying solely on the Top-N metric may not fully reflect the actual impact of each method. 
Nevertheless, these metrics are widely accepted in the FL domain, and offering an explanation is only useful if a fault location itself is correct as discussed in Section~\ref{sec:explanation}.
Therefore, this issue will not alter our conclusion that \logicfl has improved FL performance.

\section{Conclusion}
\label{sec:conclusion}

In this study, we propose \logicfl, a novel logical fault localization technique for NPEs, and present empirical evaluation results with detailed analysis.
\logicfl is designed to imitate human's deduction process of fault localization by applying logical rules to facts collected from faulty code and test execution.
In our empirical evaluation, \logicfl significantly outperforms the compared LLM-based FL techniques, and proves that a logical FL approach has great potential to provide reliable and efficient automated solutions.
Moreover, \logicfl is a lightweight tool that can run on regular laptops, significantly reducing costs compared to LLM-based techniques, which are hundreds of times more expensive. 
This demonstrates that for problems where we have well-organized knowledge, leveraging that knowledge can lead to highly effective and efficient solutions. 
While \logicfl currently focuses on NPE causes, we believe that it can be extended to address other error types, different programming languages, and even other issue types like security vulnerabilities.

\section{Data Availability}

LogicFL replication package can be found at Figshare (archived)~\cite{replication}. We also provide SWISH (SWI-prolog for SHaring) notebooks for the NPE benchmark~\cite{swish, npe_benchmark}.


\bibliographystyle{ACM-Reference-Format}
\bibliography{logicfl}

\end{document}